\documentstyle[11pt]{article}
\newcommand{\cosec}{{\rm cosec}}
\newcommand{\sech}{{\rm sech}}
\newcommand{\cosech}{{\rm cosech}}

\def\eq{\!\!\!\! &=& \!\!\!\! }
\def\br{\begin{eqnarray}}
\def\er{\end{eqnarray}}
\def\be{\begin{equation}}
\def\ee{\end{equation}}
\def\lb{\lbrack}
\def\rb{\rbrack}

\def\>{\rangle}              
\def\<{\langle}              
\def\({\left(}
\def\){\right)}
\def\[{\left[}
\def\]{\right]}

\def\j3{J_3}
\def\Ad{{\cal A}^{\dagger}}
\def\A{{\cal A}}
\def\nn{\nonumber}

\def\br{\begin{eqnarray}}
\def\er{\end{eqnarray}}
\def\be{\begin{equation}}
\def\ee{\end{equation}}
\newcommand{\sss}{\vspace{.2in}}
\newcommand{\sms}{\vspace{.1in}}
\begin{document}
~~\hfill{UICHEP-TH/01-1}\\

\vspace*{1.0in}
\centerline{\large \bf Broken Supersymmetric Shape Invariant Systems and Their Potential Algebras}
\vspace{0.1in}

\vspace{0.7in}

\begin{center}
{{\bf
   \mbox{Asim Gangopadhyaya}$^{a,}$\footnote{agangop@luc.edu, asim@uic.edu},
   \mbox{Jeffry V. Mallow}$^{a,}$\footnote{jmallow@luc.edu} and
   \mbox{Uday P. Sukhatme}$^{b,}$\footnote{sukhatme@uic.edu}
 }}
\end{center}
{\small
\noindent
a)\hspace*{.11in}
Department of Physics, Loyola University Chicago, Chicago, Illinois 60626 \\
b)\hspace*{.11in}
Department of Physics (m/c 273), University of Illinois at Chicago, Chicago, Illinois 60607 \\
}
\sss

\centerline{\bf Abstract}
\sms
Although eigenspectra of one dimensional shape invariant potentials with unbroken supersymmetry are easily obtained, this procedure is not applicable when the parameters in these potentials  correspond to broken supersymmetry, since there is no zero energy eigenstate. We describe a novel two-step shape invariance approach as well as a group theoretic potential algebra approach for solving such broken supersymmetry problems.

\newpage

In the formalism of supersymmetric quantum mechanics (SUSYQM), potentials with unbroken supersymmetry \cite{SUSY,cooper} and shape-invariance can be exactly solved by a well-known standard procedure \cite{shape_invariance}. The potential algebras of these systems have also been identified  \cite{Alhassid,Gangopadhyaya,Balantekin}, providing an alternate method of solution.

In this paper, we study shape invariant potentials in which the parameters have values such that the supersymmetry (SUSY) is spontaneously broken. For these systems, the usual shape invariance procedure does not enable one to determine the spectra, since there is no zero energy eigenstate \cite{SUSY,cooper}. Inomata and Junker \cite{Inomata} studied these systems using a  semiclassical method known as broken supersymmetric WKB (BSWKB). Their method yielded exact spectra for all solvable systems with broken supersymmetry known at the time of their work.  In ref. \cite{dutt}, these systems were solved by a judicious change of parameters, mapping them to theories with unbroken SUSY. In this paper, our approach is to provide a systematic analysis using only shape invariance. For each of the potentials considered in ref. \cite{dutt}: the three dimensional harmonic oscillator, P\"oschl-Teller I and II, we show that there exists a two-step shape invariance that renders them exactly solvable.  The first step converts the initial potential into one which can be solved in a second step using the standard procedure of unbroken SUSY. While the treatment of this paper is mathematically similar to that of ref. \cite{dutt}, the approach presented here uses  two separate forms of shape invariance and this leads in a natural manner to their underlying potential algebra. This is the first application, to our knowledge, of shape invariance to connect broken and unbroken SUSYQM potentials.

For completeness, we first provide in a brief review of supersymmetric quantum mechanics \cite{SUSY,cooper}. Then, we demonstrate the two-step shape invariance method by explicitly considering the P\"{o}schl-Teller I potential, showing the types of shape invariances it possesses, and exploiting these to determine the eigenspectrum. The same method is then extended to the three dimensional harmonic oscillator and P\"{o}schl-Teller II 
potentials.  Finally, we study the potential algebras underlying these systems, and generate their spectra by algebraic means. \sss

\noindent {\bf Supersymmetric Quantum Mechanics:}
In supersymmetric quantum mechanics  \cite{cooper}, the partner potentials $V_{\pm}(x,a_0)$ are related to the superpotential $W(x,a_0)$ by $V_{\pm}(x,a_0)=W^2(x,a_0) \pm W'(x,a_0)$,
where $a_0$ is a set of parameters. It is assumed that the superpotential is continuous and differentiable. Setting $\hbar=2m=1$, the corresponding Hamiltonians $H_{\pm}$ have a factorized form
\be \label{hpm}
H_-=\Ad \A~,~H_+=\A \Ad~,~\A \equiv {d \over {dx}} +W~,~ \Ad \equiv -{d \over {dx}} +W.
\ee
If either ${\psi_0^{(-)}}(x,a_0) \equiv \exp\left( - \int^x W(y,a_0) dy \right)$ [or $1/{\psi_0^{(-)}}(x,a_0)$] is normalizable, it is the ground state for $H_-$ [or $H_+$] and corresponds to the case of unbroken SUSY. As can be explicitly checked $\A \psi_0^{(-)} = 0$, and thus the ground state $E_0^{(-)}$ is zero. If, however, neither ${\psi_0^{(-)}}$ nor $\frac{1}{{\psi_0^{(-)}}}$ is normalizable, the SUSY is said to be spontaneously broken. The vanishing of the ground state energy is a sufficient and necessary condition for unbroken SUSY  \cite{SUSY}. In that situation, 
the Hamiltonians $H_+$ and $H_-$ have exactly the same eigenvalues, except that $H_-$ has an additional zero energy eigenstate.
More specifically, the eigenstates of $H_+$ and $H_-$ are related by
\be
\label{SUSY}
E^{(-)}_0 = 0~,~~~E^{(+)}_{n-1}\;=\;E^{(-)}_{n},~~~ \psi^{(+)}_{n-1} \propto \A \,\psi^{(-)}_{n}~~,~~~\Ad \,\psi^{(+)}_{n} \propto \psi^{(-)}_{n+1}  ~~, \quad   n=1,2, \ldots.
\ee

Supersymmetric partner potentials are called shape invariant if they both have the same $x$-dependence up to a change of parameters $a_1=f(a_0)$ and an additive constant which we denote by $R(a_0)$ \cite{cooper}.  The shape invariance condition is
\be
\label{sipv}
V_+(x,a_0) = V_-(x,a_1) + R(a_0)= V_-(x,a_1) + g(a_1)- g(a_0)~,
\ee
where, for future convenience we have expressed the constant $R(a_0)$ as a difference $g(a_1)- g(a_0)$. The property of shape invariance, coupled with unbroken SUSY, permits an immediate analytic determination of energy eigenvalues and eigenfunctions   \cite{shape_invariance,dutt}, of $H_-(x,a_0)$. The ground state energy of $H_-(x,a_0)$ is zero, as we have assumed unbroken SUSY. If the change of parameters $a_0 \rightarrow  a_1$ does not break SUSY, $H_-(x,a_1)$ then also has a zero energy ground state, and the corresponding eigenfunction is given by $\psi_0^{(-)}(x, a_1) \propto \exp \(- \int^x_{x_0} W(y,a_1) dy \)$. Now using eqs. (\ref{SUSY}) and (\ref{sipv}) we have
\be
E_1^{(-)} = R(a_0)~,~~ \psi_1^{(-)}(x,a_0) = \Ad(x,a_0)\, \psi_0^{(+)}(x,a_0)=
\Ad(x,a_0) \,\psi_0^{(-)}(x,a_1)~.
\ee
Thus for unbroken SUSY, the eigenstates of the potential $H_-(x,a_0)$ are
\br \label{eq4}
E_0^{(-)} \eq 0~,~E_n^{(-)}=\sum_{k=0}^{n-1}R(a_k)=\sum_{k=0}^{n-1} \[g(a_{k+1})- g(a_{k})\]=g(a_n)- g(a_0)~,\\
\psi_0^{(-)} &\!\!\!\!\propto&\!\!\!\! e^{- \int^x_{x_0} W(y,a_0) dy}~,~
\psi_n^{(-)}(x,a_0)=\left[-\frac{d}{dx}+W(x,a_0)\right]
\psi_{n-1}^{(-)}(x,a_1)~,~~(n=1,2,3,\ldots)~.\nonumber
\er
Consequently, for a given shape invariant Hamiltonian $H_-(x,a_0)$, each of its eigenvalues $E_n^{(-)}(a_0)$ and eigenfunctions  $\psi_n^{(-)}(x,a_0)$ is obtainable from a related Hamiltonian $H_-(x,a_n)$. However, these formulas are only valid provided that the Hamiltonians retain unbroken SUSY under change of parameters $a_{k+1} = f(a_k)$, $k=1...n$.  In previous work on shape invariant potentials, changes of parameters corresponding to  translation $a_1 = a_0 + \beta$ \cite{dutt} and scaling $a_1 = q a_0$ with $0 <q \le 1$ \cite{Barclay} have been discussed. However, a reflection change of parameters $a_1 = -a_0$, even if it maintained shape invariance, was not seriously considered since it could not maintain unbroken SUSY for the hierarchy of potentials built on $H_-$.\sss

\noindent{\bf Shape Invariant Potentials with Broken SUSY:}
We now develop and describe our two step approach for solving shape invariant problems with broken SUSY.  To illustrate this approach, it is best to consider a specific example. Consider the P\"{o}schl-Teller I superpotential 
\be \label{wpt1}
W(x,A,B) = A \tan x - B \cot x~~;~~~~0<x<\pi/2~~.
\ee
The supersymmetric partner potentials are given by
\br \label{vpt1}
V_-(x,A,B) \eq  A(A-1) \sec^2x +B(B-1) \cosec^2x - (A+B)^2~~;
\nn \\
V_+(x,A,B) \eq A(A+1) \sec^2x +B(B+1) \cosec^2x - (A+B)^2~~.
\er
The function $\psi_0^{(-)}(x,A,B) \equiv \exp\(  - \int^x 
W(y,A,B)~dy  \)$ is given by 
$\psi_0^{(-)}(x,A,B) =  \cos^{A}x\sin^{B}x$.
For $A>0, B>0$, the function $\psi_0^{(-)}(x,A,B)$ is normalizable and proportional to the ground state wave function, implying unbroken SUSY. Similarly, for $A<0, B<0$, the function $1/\psi_0^{(-)}(x,A,B)$ is normalizable, and one again has unbroken SUSY, and the spectrum is obtainable from shape invariance. However, for $A>0, B<0$,
$\psi_0^{(-)}(x,A,B)$ is not normalizable due to its divergent behavior at $x=0$ while its reciprocal $\frac{1}{\psi_0^{(-)}}$ is not normalizable due to its divergent behavior at $x=\frac{\pi}{2}$, and hence SUSY is broken. Likewise, for $A<0, B>0$, one again has a broken SUSY situation. As mentioned before, the broken SUSY case has no zero energy state, and the standard shape invariance procedure [eq. (\ref{SUSY})] cannot be used to get the eigenstates.

Let us now focus on the case $A>0, B<0$. The eigenstates of $V_\pm(x,A,B)$ are related by
\be \label{relations_bsusy}
\psi_n^{(+)}(x,a_0) =  A(x,a_0)\psi_n^{(-)}(x,a_0) ~;~ \psi_n^{(-)}(x,a_0) =  \Ad(x,a_0)\psi_n^{(+)}(x,a_0) ~;~
E_n^{(-)}(a_0) = E_n^{(+)}(a_0) ~.
\ee

The potentials of eq.~(\ref{vpt1}) are shape invariant. In fact there are two possible relations between parameters such that these two potentials exhibit shape invariance. One of them is the conventional $(A \rightarrow A+1,~~ B \rightarrow B+1)$.  The shape invariance condition is given by 
\be\label{sic1}
V_+(x,A,B) = V_-(x,A+1,B+1) + (A+B+2)^2 - (A+B)^2 ~~. 
\ee
For $B$ sufficiently large and negative, $B+1$ is also negative; thus the superpotential resulting from this change of parameters still falls in the broken SUSY category, and hence $E_0^{(-)}(a_0) \neq 0$. In the absence of this crucial result, even with shape invariance, we are not able to proceed further. This is the reason why the spectra of these systems remained undetermined by the methods of SUSYQM for so long. 

The second possibility is $(A \rightarrow A+1,~~ B \rightarrow -B)$. The corresponding relationship is given by
\be\label{sic2}
V_+(x,A,B) = V_-(x,A+1,-B) + (A-B+1)^2 - (A+B)^2 . 
\ee
This change of parameters $(A \rightarrow A+1,~~ B \rightarrow -B)$ leads to a system with unbroken SUSY, since the parameter $B$ changes sign. Hence the ground state of the system with potential $V_-(x,A+1,-B)$ is guaranteed to be at zero energy. {}From eq. (\ref{sic2}), we see that potentials $V_+(x,A,B)$ and $V_-(x,A+1,-B)$ differ only by a constant, hence
$$\psi_+(x,A,B) = \psi_-(x,A+1,-B)~~;~~
E_n^{(+)}(A,B) = E_n^{(-)}(A+1,-B) + (A+1-B)^2 - (A+B)^2 ~~. $$
Thus, if we knew the spectrum of the unbroken SUSY $H_-(x,A+1,-B)$, we would be able to determine the spectrum of $H_+(x,A,B)$ with broken SUSY. Since the parameters of the potential $V_-(x,A+1,-B)$ lie in the region necessary for unbroken SUSY, there is an extensive machinery, already at hand, to determine the eigenstates of this potential \cite{cooper}. The results are
\begin{equation}\label{eq12}
E_n^{(-)}(A+1,-B) = (A+1-B+2n)^2 - (A+1-B)^2. 
\end{equation}
When combined with eqs. (\ref{relations_bsusy}) and (\ref{sic2}), one gets
\br\label{eq13}
E_n^{(-)}(A,B) \eq (A+1-B+2n)^2 - (A+B)^2~;~\nonumber\\
\psi_n^{(-)}(y,A,B)\eq(1+y)^{(1-A)/2}
(1-y)^{B/2}P_n^{(B-1/2,1/2-A)}(y),
\er
where $y=\cos(2x)$ and $P_n^{(\alpha,\beta)}(y)$ are Jacobi polynomials, in agreement with the results of ref. \cite{dutt}. 

We have considered the P\"{o}schl-Teller I potential in detail. Very similar analyses can be used for determining the eigenvalues and eigenstates of the three dimensional harmonic oscillator, as well as the P\"{o}schl-Teller II problem. We will now briefly describe the treatment of these potentials.

The problem of the three dimensional harmonic oscillator with broken SUSY is described by the superpotential
\begin{equation}
\label{wr}
W(r,l,\omega)=   \frac{1}{2} ~\omega r -\frac{l+1}{r}~~;\quad l<-1 ~.
\end{equation}
The function $\psi_0^{(-)}(r,l,\omega) = \exp\(  - \int^r W(r',l,\omega)~dr'  \) = r^{l+1} e^{ - \omega r^2}$ diverges near $r \rightarrow 0$ for $l<-1$ and hence corresponds to the case of broken SUSY. The supersymmetric partner potentials are
\be
\label{potentials_H}
V_+(r,l,\omega)= \frac{\omega^2 r^2}{4}  +\frac{(l+1)(l+2)}{r^2} - \left(l+\frac{1}{2} \right)\omega,~
V_-(r,l,\omega)= \frac{\omega^2 r^2}{4} +\frac{l(l+1)}{r^2} - \left(l+\frac{3}{2} \right)\omega.
\ee
These two partner potentials are shape invariant since 
\be\label{sic1_H}
V_+(r,l,\omega)=V_-(r,l+1,\omega) + 2\omega~. 
\ee
For sufficiently large negative values of $l$, the potential $V_-(r,l+1,\omega)$ also lies in the realm of broken SUSY. Thus, usual SUSYQM method again fails to deliver the spectrum.
However, there is another change of parameters that also maintains shape invariance between these two partner potentials, namely
\be\label{sic2_H}
V_+(r,l,\omega)= V_-(r,-l-2,\omega) -(2l+1)\omega~~.
\ee
Since $l<-1$, the potential $V_-(r,-l-2,\omega)$ has unbroken SUSY and has a zero energy ground state. Combining the results of eqs. (\ref{potentials_H}), (\ref{sic1_H}) and (\ref{sic2_H}), we get
\be
E^{(+)}_n(l,\omega)= (2n-2l-1)\omega~.
\ee
Our last example is the P\"{o}schl-Teller II potential described by
\be \label{wpt2}
W(r,A,B) = A \tanh r - B \coth r~~;~~~~0<r<\infty~~.
\ee
The function $\psi_0^{(-)}(r,A,B) \equiv \exp\(  - \int^r 
W(r',A,B)~dr'  \)$ is given by 
$\psi_0^{(-)}(r,A,B) =  \cosh^{-A}r \, \sinh^{B}r$.
Here, for $A>0$ and $B<0$,  neither $\psi_0^{(-)}(r,A,B)$ nor its inverse are normalizable, and hence we have a system with broken SUSY. The supersymmetric partner potentials are given by
\br
V_+(r,A,B) \eq -A(A-1) \sech^2r +B(B-1) \cosech^2r + (A+B)^2 \nn \\
V_-(r,A,B) \eq -A(A+1) \sech^2r +B(B+1) \cosech^2r + (A+B)^2~~.
\er
Here too we have two possible relations  between parameters for these potentials to be shape invariant. They are 
\be\label{sic1_Pt2}
 {V}_+(r,A,B) =  {V}_-(r,A-1,B-1)+(A+B)^2 - (A+B-2)^2~~,
\ee
and
\be\label{sic1_Pt2}
 {V}_+(r,A,B) =  {V}_-(r,A-1,-B)+(A+B)^2 - (A-B-1)^2~~.
\ee
As explained in the previous two examples, the first transformation does not lead to the determination of the spectrum since the  new parameters 
$(A-1,~B-1)$ still correspond to broken SUSY for sufficiently large positive value of $A$. However, in the second type of transformation, the new values of the parameters $(A-1,~-B)$ lie in the domain of unbroken SUSY and hence the spectrum of ${V}_-(r,A-1,-B)$ can be determined by standard methods of supersymmetric quantum mechanics. The resulting spectrum is given by
$$ E_n^{(-)}(A,B)=(A+B)^2 - (A-1-B-2n)^2.$$\sss

\noindent{\bf Spectra of Broken SUSY Problems using Potential Algebras:}
The shape invariance based approach to broken supersymmetric potentials discussed above also naturally leads us to the underlying potential algebra, which in turn allows us to determine the spectrum by algebraic means\cite{Gangopadhyaya,Balantekin}. We will see below that the algebraic approach closely mimics the SUSYQM approach, and, analogous to the combination of the two step invariances, here one has to combine two different algebras.

Again, we consider the P\"oschl-Teller I potential to exemplify the algebraic method. Let us first examine the unbroken sector of the theory \cite{Gangopadhyaya,Gango_CJP}. Consider a set of three operators \cite{dutt_PRA_99}
\be
\label{cc1}
    J^+ =  c^{-1}\,\Ad\(x, \alpha-N, \beta-N\) ~~,~~
    J^-=  \A\(x,\alpha-N, \beta-N\) c~~,~~    J_3 = N \equiv c^{\dagger} c~~,
\ee
where $\A \(x,\alpha-N, \beta-{N}\)$ is obtained by replacing the parameters $A$ and $B$ by the operators $\alpha-N$ and $\beta-{N}$ in eqs. (\ref{wpt1}) and (\ref{hpm}). $\alpha$ and $\beta$ are constants to be determined later. Operators  $c, c^{\dagger}$ and $c^{-1} $ are chosen so as to 
satisfy\footnote{A coordinate realization of such operators is given by $c = e^{i\phi}, ~c^{-1}=e^{-i\phi}$ and $c^{\dagger} = i\frac{\partial}{\partial \phi}\, e^{-i\phi}$, where $\phi$ is some arbitrary real variable.}
$[c,c^{\dagger}]=1$, and $ c\, c^{-1} = c^{-1}\, c = 1$. It follows that for any Taylor expandable function $f(N)$, one has $ f(N)\,c = c\,f(N-1)$ \cite{dutt_PRA_99}.
One finds that the operators $J_3,J^{\pm}$ satisfy
\be
\label{pot-alg1}
    \lb J_3,J^{\pm} \rb = \pm J^{\pm} ~;~~~
    ~\lb J^{+}, J^{-} \rb = (\alpha+\beta-2N)^2-(\alpha+\beta-2N+2)^2~.
\ee 
The last commutation relation is a consequence of the shape invariance condition 
(\ref{sic1}), where the Hamiltonians corresponding to the P\"oschl-Teller I potential are given by
$\widetilde{H}_+\,\(x,\alpha-N, \beta-N\)=J^{-}J^{+} $ and 
$\widetilde{H}_-\,\(x,\alpha-N+1,\beta-N+1\)=J^{+}J^{-} $. 
To determine the spectrum of these Hamiltonians, we define,
\be\label{representation}
J^+ |n\> = a_{n+1} |n+1\>~;~~~J^- |n\> = a_{n} |n-1\>~~.
\ee
Applying the commutator $\lb J^{+}, J^{-} \rb$ of eq. (\ref{pot-alg1}) on state $| n \>$, one gets: 
\br \label{gory}
a_{n+1}^2-a_n^2 = (\alpha+\beta+2-2N)^2-(\alpha+\beta-2N)^2 \equiv D(N-1)-D(N)~~;\nn \er
and its iteration yields
\be
a_1^2-a_{0}^2 = D(-1)-D(0)~,~
a_2^2-a_{1}^2 = D(0)-D(1)~,\ldots,
a_{k}^2-a_{k-1}^2 = D(k-2)-D(k-1) ~.
\ee
Adding all the rows of eq. (\ref{gory}), and demanding $a_0=0$ (this follows from  $J^- |0\> = 0$ for unbroken SUSY) one gets $ a_{k}^2= D(-1)-D(k-1)= (\alpha+\beta+2)^2-(\alpha+\beta+2-2k)^2$. This gives
\br
\widetilde{H}_-\,\(x,\alpha-k+1,\beta-k+1\)|k\> = \left[ (\alpha+\beta+2)^2-(\alpha+\beta+2-2k)^2 \right] |k\> \nn 
\er
Now, identifying $\alpha-k+1=A$ and $\beta-k+1=B$ in the above Hamiltonian yields
\br
\widetilde{H}_-\,\(x,A,B \)|k\> = \left[ (A+B+2k)^2-(A+B)^2 \right] |k\> ~~,k=0,1,2,\ldots~,\nn 
\er
in agreement with eq. (\ref{eq12}).

 Now, for the broken SUSY sector, we consider three additional operators
\br
\label{cc2}
    K^+ =  c^{-1}\,\Ad\(x, \gamma-N, (-1)^{N}\delta\) ~,~
    K^-=  \A\(x,\gamma-N, (-1)^{N}\delta\) c~,~
    K_3 = N ~,
\er
with their algebra given by 
\br
\label{pot-alg2}
    \lb K_3,K^{\pm} \rb = \pm K^{\pm} ~,~    ~\lb K^{+}, K^{-} \rb \eq (\gamma-N+(-1)^{N}\delta)^2-(\gamma-N - (-1)^{N}\delta+1)^2~.~~~
\er
The last commutation relation follows from the shape invariance condition 
(\ref{sic2}) and $K^{-}K^{+}= H_+\,\(x,\gamma-N, (-1)^{N}\delta\)$ and 
$K^{+}K^{-}= H_-\,\(x,\gamma-N+1, -(-1)^{N}\delta\)$.
As before, we define
\be\label{representation2}
K^+ |n\> = b_{n+1} |n+1\>~;~~~K^- |n\> = b_{n} |n-1\>~~,
\ee
and one gets
\br 
\label{gory2}
b_n^2-b_{n+1}^2 = (\gamma-n+(-1)^{n}\delta)^2- (\gamma-n+1 - (-1)^{n}\delta)^2~~.\nn 
\er
At this point we would like to choose the starting values of $n$ and the constants $\gamma,~\delta$ to be such that ($\gamma-n$), and $(-1)^{n}\delta$ fall in the broken SUSY domain, where $n$ is an eigenvalue of the operator $N$ associated with a state $|n\>$.  In algebraic terms, this implies that unlike the unbroken SUSY case, we cannot determine the values of $b_n^2$  as we do not know any state that is annhilated either by $K_-$ or by $K_+$, and hence we can not set $b_0=0$. All we can do is to relate eigenstates of the Hamiltonians 
$K^{-}K^{+}$ and $K^{+}K^{-}$:
\br\label{temp}
\psi_n^{(+)}\(x,\gamma-n, (-1)^{N}\delta\) \eq \psi_n^{(-)}\(x,\gamma-n+1, -(-1)^{N}\delta\)~,\nn \\
E_n^{(+)} \(\gamma-n, (-1)^{N}\delta\) \eq E_n^{(-)} \(\gamma-n+1, -(-1)^{N}\delta\)+b_n^2-
b_{n+1}^2~.
\er
However, due to the inversion of the second parameter $(-1)^{N}\delta$ in eqs. (\ref{cc2}) and (\ref{pot-alg2}),   the Hamiltonian $ K^{+}K^{-} = H_-\,\(x,\gamma-N+1, -(-1)^{N}\delta\)$ corresponds to unbroken SUSY, and the discussion above can now be used to determine its eigenstates. This discovery lets us join the algebras of $K^\pm,~K_3$ with that of $J^\pm,~J_3$.  We demand that for a specific eigenvalue of $N$, 
 $$H_-\,\(x,\gamma-N+1, -(-1)^{N}\delta\)=\widetilde{H}_-\,\(x,\alpha-N+1,\beta-N+1\),$$
i.e. $\alpha=\gamma$ and $\beta=N-1-(-1)^{N}\delta$. Thus, the energy of the Hamiltonian $H_-\,\(x,\gamma-N+1, -(-1)^{N}\delta\)$, with $\gamma-N+1$ and 
$-(-1)^{N}\delta$ as its parameters, is given by 
\br
E_k^{(+)} \(\gamma-k, (-1)^{k}\delta\) \eq E_k^{(-)} \(\gamma-k+1, -(-1)^{k}\delta\)+b_{k+1}^2-
b_{k}^2~ \nn \\
\eq  
\( \gamma-k+1-(-1)^{k}\delta+2k \)^2 - \( \gamma-k+(-1)^{k}\delta \)^2 
\nn
\er
Thus, Hamiltonians 
$H_\pm\,\(x,A,B\)$, with $A,~B$ in the broken SUSY range, have eigenenergies 
\be
E_k^{(\pm)}(A,B) = \left[ \( A+1-B+2k \)^2 - \( A+B \)^2 \right]~,
\ee
as obtained before.

A similar procedure yields the eigenstates of the P\"{o}schl-Teller II and harmonic oscillator  potentials in the broken SUSY phase.

We thank the U.S. Department of Energy for partial financial support.

\end{document}